\documentclass[pra,aps,onecolumn,english,superscriptaddress]{revtex4-1}
\usepackage[T1]{fontenc}
\usepackage[latin9]{inputenc}
\usepackage{xcolor}
\usepackage{amsmath}
\usepackage{amsthm}
\usepackage{amssymb}
\usepackage{graphicx}
\usepackage{subfigure}
\usepackage{braket}
\DeclareMathOperator{\Tr}{Tr}

\usepackage[strict]{changepage}

\makeatother

\usepackage{babel}

\begin{document}
\title{Multilevel quantum thermodynamic swap engines}

\author{Massimiliano F. Sacchi}
\email{msacchi@unipv.it}
\affiliation{CNR - Istituto di
  Fotonica e Nanotecnologie, Piazza Leonardo da Vinci 32, I-20133,
  Milano, Italy,}  \affiliation{QUIT Group, Dipartimento di Fisica,
  Universit\`a di Pavia, via A. Bassi 6, I-27100 Pavia, Italy.}

\begin{abstract}
  We study energetic exchanges and fluctuations in two-stroke quantum
  thermodynamic engines where the working fluid is represented by two
  multilevel quantum systems, i.e. qudits, the heat flow is allowed by
  relaxation with two thermal reservoirs at different temperatures,
  and the work exchange is operated by a partial-swap unitary
  interaction. We identify three regimes of operation (heat engine,
  refrigerator, and thermal accelerator), present the thermodynamic
  uncertainty relations between the entropy production and the
  signal-to-noise ratio of work and heat, and derive the full joint
  probability of the stochastic work and heat. Our results bridge the
  gap between two-qubit and two-mode bosonic swap engines, and show which
  properties are maintained (e.g., a non fluctuating Otto efficiency)
  and which are lost for increasing dimension (e.g., small violations
  of the standard thermodynamic uncertainty relations or the
  possibility of beating the Curzon-Ahlborn efficiency).
\end{abstract}

\maketitle
\section{Introduction}
A growing interest has been recently devoted to thermodynamic engines
where the working systems are operated at the nanoscale
\cite{nano1,nano2,nano3}, from electronic devices \cite{ventra,soth}
to biological or chemical systems \cite{gnes,rit,rao}.  Differently
from the usual scenery of macroscopic thermodynamics, when the working
substances are elementary and small enough and generally when the
discrete nature of the energy spectrum is relevant, the fluctuations
of the thermodynamical variables become very important, and the
theoretical approach in terms of stochastic thermodynamics
\cite{stoc1,stoc2,stoc3} turns out to be very useful to address the
problem of quantifying the efficiency along with the stability and
reliability of quantum thermodynamic engines.
 
\par In fact, fluctuation relations
\cite{evans,gal,jar97,crook,piecho,jarz3,seif2,marc,saito,tch,
  andrie,stoc2,esp2,cth,sini,camp,stoc3, frq,hang,esp3} pose rigid
constraints on the statistics of heat, work and entropy production in
terms of the symmetries of the elemental microscopic dynamics.  More
recently, so-called thermodynamic uncertainty relations (TUR) have
been developed \cite{bar,pietz,ging2,pole,pietz2,horo,proes2,agar,
  koy,bar2,brad,piet,holu,macie,Li,sary,dech,proes,bar3,guar,ging},
where the signal-to-noise ratio of observed work and heat has been
related to the entropy production.  For example, these TURs balance
the tradeoff between entropy production and fluctuations of output
power, namely the precision of a heat engine, such that working
systems operating at vanishing entropy production entail a divergence
in the relative output power fluctuations.  Fluctuation relations and
TURs have been independently developed, but lately they have been
connected within various approaches and operational assumptions
\cite{proes2,vanvu,potts,vanvu2,timpa,zhang,vanvu3,merh,gin3,salaz},
suited also to the analysis of quantum thermodynamic stroke engines
\cite{timpa,miobos}.
 
\par One of the most studied quantum thermodynamic engines is based on
the Otto cycle
\cite{otto4,ot1,ot2,ot3,ot4,ot5,cpf,dec,ot6,strob,miobos,picc}, with
possible implementations by different physical systems as working
fluid, e.g.  ion traps, Cooper-pair boxes, or quantized modes of the
radiation field. The advantage of stroke Otto engines is that a deep
study is amenable, since heat and work strokes are clearly separated:
in a part of the dynamics, the systems are coupled to thermal
reservoirs and are allowed to relax with fixed Hamiltonian by heat
exchange, while in a different part they are isolated from the baths
and work is externally supplied or extracted by driven Hamiltonian. In
fact, a thorough derivation of the full stochastic heat and work
distribution can be accomplished when the working strokes are  operated
by a partial-swap unitary interaction, when the working fluid is
represented by two bosonic modes or two qubits \cite{miobos}.

\par Building on the approach of Ref. \cite{miobos}, in this paper we
study a two-stroke thermodynamic engine where the working fluid is
abstractly given by two multilevel quantum systems (i.e. qudits), each
with equally-spaced energy levels. These are alternately coupled to
their own thermal bath at different temperatures allowing heat
exchange, whereas the working stroke is implemented by a unitary
interaction with tunable partial swap in order to extract or supply
work. In the situation of perfect swap operation, this model
interpolates the case of two qubits with that of two harmonic
oscillators under $50/50$ frequency conversion \cite{miobos}, and may
find application with different high-dimensional quantum systems, as
Rydberg atoms \cite{ryd}, polar molecules \cite{pol}, trapped ions
\cite{tra}, NMR systems \cite{nmr0}, cold atomic ensembles
\cite{cold1,cold2}, and discretized degrees of freedom of photons
\cite{zeil}.

\par By the joint estimation of work and heat via a
two-point-measurement scheme \cite{stoc3,der,th,camp}, we obtain the
characteristic function that provides all moments of work and
heat. Three regimes of operation are identified, where the periodic
protocol works as a heat engine, a refrigerator, or a thermal
accelerator. The model is shown to achieve the Otto efficiency,
independently of the dimension, the temperature of the reservoirs, and
the coupling parameter.  We present an exact relation between the
signal-to-noise ratio of work and heat and the average entropy
production of the engine, thus linking together average extracted
work, fluctuations, and entropy production. From these relations we
derive thermodynamic uncertainty relations that are satisfied in all
the regimes of operations and for any dimension. Similarly to the case
of two-qubit stroke engines \cite{timpa,miobos}, a small violation of
the standard TUR is observed, which, however, is rapidly washed out
for increasing dimension of the working qudits or for decreasing
coupling strength. A bound of the
efficiency in terms of the first two moments of the work distribution
is also obtained.  Finally, we provide the full joint probability of
the discrete stochastic work and heat in closed form, which allows to
explicitly verify a detailed fluctuation theorem. We conclude the
paper with a preliminary analysis of the finite-time cycle, i.e. by
considering partial thermalization strokes, and study the resulting
output power in the case of perfect-swap unitary interaction. In this
case we show the possibility of beating the Curzon-Ahlborn efficiency
\cite{ca1,ca4,ca2}, which however is reduced for increasing dimension.

\section{A two-qudit swap engine}
Throughout the paper we fix natural units for both Planck and
Boltzmann constants, namely $\hbar =k_B=1$. The thermodynamic engine
under investigation is based on a working fluid given by
a couple of qudits $A$ and $B$, i.e. two $d$-level quantum systems, 
each one with equally-spaced energy levels. We can write their free Hamiltonian as
\begin{eqnarray}
H_C=\omega _C \sum_{n=0}^{d-1} n |n \rangle 
\;,
\end{eqnarray}
with $C=A,B$. 
Initially, the two qudits  are in thermal equilibrium with
their own ideal bath at temperature $T_A$ and $T_B$, respectively, and we fix $T_A >
T_B$. Hence, the initial state is described by the tensor product
of Gibbs thermal states, i.e. 
\begin{eqnarray}
  \rho _0 = \frac{e ^{-\beta _A H_A}}{Z_A} \otimes \frac{e ^{-\beta _B H_B}}{Z_B} 
  \;,\label{eqq}
\end{eqnarray}
with $\beta _X = 1/T_X$ and $Z_X=\Tr [e^{-\beta _X
    H_X}]=\frac{1-e^{- d\beta _X \omega _X}}{1-e^{- \beta _X \omega _X}}$. The two
qudits are then isolated from their thermal baths and are allowed to
interact in a  time window $[0,\tau _w]$ via the time-dependent interaction
\begin{eqnarray}
  H(t)=  \kappa \,e^{-i (H_A+H_B)t} E e^{i (H_A+H_B)t}\label{ht}  
\;, 
\end{eqnarray}
where $\kappa $ is a real coupling parameter and
$E$ denotes the swap operator, which acts on two-qudit
states as $E |\psi \rangle \otimes |\varphi \rangle = |\varphi \rangle
\otimes |\psi \rangle $. 
In the interaction picture where
\begin{eqnarray}
\rho _I (t)=U_0^\dag (t) \rho (t) U_0 (t)
\;,
\end{eqnarray}
with $U_0 (t) =\exp [-i (H_A +H_B)t]$, the interaction Hamiltonian is
clearly 
\begin{eqnarray}
H_I (t) \equiv H_ I = H(0)=\kappa\, E
  \;,
\end{eqnarray}
and hence one has $\rho _I (t)= e^{-i H_I t} \rho _I (0)e^{i H_I t} $. Going
back to the Schr\"odinger picture we obtain the evolution
\begin{eqnarray}
  \rho (t) \equiv U(t) \rho (0) U^\dag (t)
  =U_0 (t) e^{-i H_I t} \rho (0) e^{i H_I t} U_0 ^\dag (t)
  \;.
\end{eqnarray}
Since $E^2 =I$ we can also write
\begin{eqnarray}
V_\theta \equiv e^{-i H_I \tau _w}= \cos \theta I -i \sin \theta E\;, 
\end{eqnarray}
with $\theta = \kappa \tau _w$, and hence $U(\tau _w)= U_0 (\tau _w)
V_\theta $. 

\par After the interaction the two qudits are reset to their
equilibrium state of Eq. (\ref{eqq}) via complete thermalization by
their respective baths.  The procedure can be sequentially repeated
and leads to a two-stroke engine. For each cycle the energy change
in qudit $A$ corresponds to the heat $Q_H$ released by the hot bath,
i.e. $Q_H = -\Delta E_A$, and similarly for qudit $B$ we have $Q_C=
-\Delta E_B$, corresponding to the heat dumped into the cold reservoir
(heat is positive when flowing out of a reservoir).  The work W is
supplied ($W>0$) or extracted ($W<0$) during the unitary interaction,
and the first law $W=-Q_H -Q_C = \Delta E_A +\Delta E_B$ holds. The
entropy production per cycle is then given by $\Sigma = -\beta _A Q_H
-\beta _B Q_C = (\beta _B -\beta _A)Q_H +\beta _B W$.

\par We characterize the engine by the independent
variables $W$ and $Q_H$, and consider the characteristic function
$\chi(\lambda , \mu )$,  where $\lambda$ and $\mu$ denote the
counting parameters for work and
heat, so that all moments can be recovered as
\begin{eqnarray}
  \langle W^l Q_H ^s \rangle = (-i)^{l+s}\left.
\frac{\partial  ^{l+s} \chi (\lambda ,\mu )}{\partial \lambda ^l
    \partial \mu ^s}\right
  |_{\lambda=\mu =0}
\label{mom}\;.
\end{eqnarray}

By adopting the two-point measurement protocol
\cite{stoc3,der,th,camp} typically considered in the derivation of
Jarzynski equality \cite{j97} to jointly estimate $W$ and $Q_H$, in
the present scenario the characteristic function can be written as
\begin{eqnarray}
\chi (\lambda ,\mu )=\Tr [U^\dag (\tau _w) (e^{i (\lambda -\mu) H_A}
  \otimes e^{i \lambda H_B})
    U (\tau _w) (e^{-i (\lambda - \mu )H_A}\otimes e^{- i \lambda H_B})
    \rho_0 ]
\;.\label{chi1}
\end{eqnarray}
Equation (\ref{chi1}) can be obtained along similar lines as in
Refs. \cite{cth,camp}, and a brief derivation is given in Appendix A
for the sake of the reader.
The explicit evaluation of the characteristic function is presented in
Appendix B with the following result
\begin{eqnarray}
  \chi (\lambda ,\mu )
= \cos^2 \theta + \sin^2 \theta \frac{\sinh \left ( \frac{\beta _A
    \omega _A}{2}\right )\sinh \left ( \frac{\beta _B
    \omega _B}{2}\right )\sinh \left [ \frac d2 (\beta _A
    \omega _A + i \xi  ) \right ]\sinh \left [ \frac d2 (\beta _B
    \omega _B - i \xi  ) \right ]}
{\sinh \left ( \frac{ d \beta _A
    \omega _A}{2}\right )\sinh \left (  \frac{d \beta _B
    \omega _B}{2}\right )\sinh \left [ \frac 12 (\beta _A
    \omega _A + i \xi  ) \right ]\sinh \left [ \frac 12 (\beta _B
    \omega _B - i \xi  ) \right ]}
  \;,\label{chif}
\end{eqnarray}
where $\xi = (\omega _A -\omega _B) \lambda - \omega _A \mu$.  We note
the identity $\chi [i\beta _B , i(\beta _B -\beta _A) ]=1$,
corresponding to the standard fluctuation theorem $\langle e^{-\Sigma
}\rangle =1$. In fact, the
stronger relation $ \chi [i\beta _B - \lambda , i(\beta _B -\beta _A)
  -\mu ]=\chi ( \lambda , \mu )$ holds \cite{nota}, which is equivalent to the detailed
fluctuation theorem \cite{andrie,cth,sini,frq}
\begin{eqnarray}
\frac{p(W,Q_H)}{p(-W,-Q_H)}= e^{(\beta _B- \beta _A )Q_H +\beta _B
  W}=e^ \Sigma\;.\label{dett}
\end{eqnarray}
We notice that for $\theta =\pi/2 $ the unitary $V_{\pi/2}$ performs a
swap gate which exchanges the states of the two quantum systems. In
this situation, in the limit $d \rightarrow \infty$ the two-qudit
model recovers the two-mode bosonic Otto engine of Ref. \cite{miobos}
under $50/50$ frequency conversion \cite{notafc}.  \par Since $\chi
(\lambda ,\mu) $ is a function of the single variable $\xi$, one has
$\partial _\mu \chi = \frac{\omega _A}{\omega _B -\omega _A} \partial
_\lambda \chi $. Hence, from Eq. (\ref{mom}) one obtains the symmetry
relations
\begin{eqnarray}
  \langle W^l Q_H ^s \rangle
=
     \left ( \frac {\omega _A} {\omega _B -\omega
       _A}\right ) ^s \langle W^{l+s} \rangle  
\;,\label{sym}
\end{eqnarray}
and, from the first law, 
$\langle Q_C ^l \rangle = (- \omega
_B /  \omega _A)^l \langle Q_H ^l \rangle $. It follows that the
average entropy production per cycle is given by
\begin{eqnarray}
  \langle \Sigma \rangle
  \equiv - \beta _A  \langle Q_H \rangle  -
  \beta _B \langle Q_C \rangle =
   \frac{\beta _A \omega
     _A - \beta_ B \omega _B }{\omega _A -
   \omega _B} \langle W \rangle \;.\label{entp} 
\end{eqnarray}
From the characteristic function we can now evaluate the average work
per cycle, which is given by  
\begin{eqnarray}
  \langle W \rangle =\frac {\sin^2\theta }{2} (\omega _B -\omega _A)
  \left \{     \coth ( \beta _A \omega _A /2)  -d  \coth ( d\beta _A
  \omega _A /2)
   -[\coth ( \beta _B \omega _B /2)-d \coth ( d\beta _B \omega _B
      /2)] \right \}
    \;.\label{ww}
\end{eqnarray}
Correspondingly, from Eq. (\ref{sym}), the heat exchanged with the hot
reservoir is
$  \langle Q_H \rangle = \frac {\omega _A}{\omega _B -\omega _A}
\langle W \rangle $. 
We note that for any positive integer $d$ the function $\coth (x/2)- d \coth (d
x/2)$ is  monotonic decreasing versus $x$. Hence, we can
identify three regimes of operation of the quantum thermodynamic 
machine, namely
\begin{eqnarray}
  a)&&\ \ \ \ 1  < \frac{\omega _A} {\omega _B} < \frac{T_A}{T_B}
  \qquad\mbox{heat engine},
  \nonumber \\ b)&&\ \ \ \  
  \frac {\omega _A}{\omega _B} > \frac{T_A}{T_B} \qquad \ \ \ \ \ \mbox{refrigerator},
  \nonumber \\c)&&\ \ \ \ 
\frac {\omega _A}{\omega _B} < 1  \qquad \ \ \ \ \ \ \ \ 
\mbox{thermal accelerator},
\nonumber 
\end{eqnarray}
where we have respectively
\begin{eqnarray}
a)&&\ \ \ \ \langle W \rangle <0,\qquad  \langle Q_H 
\rangle >0,\qquad \langle Q_C \rangle <0 \,;\nonumber \\
b)&&\ \ \ \ \langle W \rangle > 0,\qquad  \langle Q_H \rangle <0,
\qquad \langle Q_C \rangle > 0 \,; \nonumber \\
c)&&\ \ \ \ \langle W \rangle >0,\qquad  \langle Q_H \rangle >0,
\qquad \langle Q_C \rangle <0\,.
\nonumber
\end{eqnarray}
In Fig. 1 we plot the average work, heat and entropy production for
dimensions $d=2,4,$ and 8, for parameters
$T_A=1,\ T_B=2,\ \theta=\pi/2,\ \omega_A=1$ versus $\omega _B/\omega
_A$.
\begin{figure}[ht]
  \includegraphics[scale=0.45]{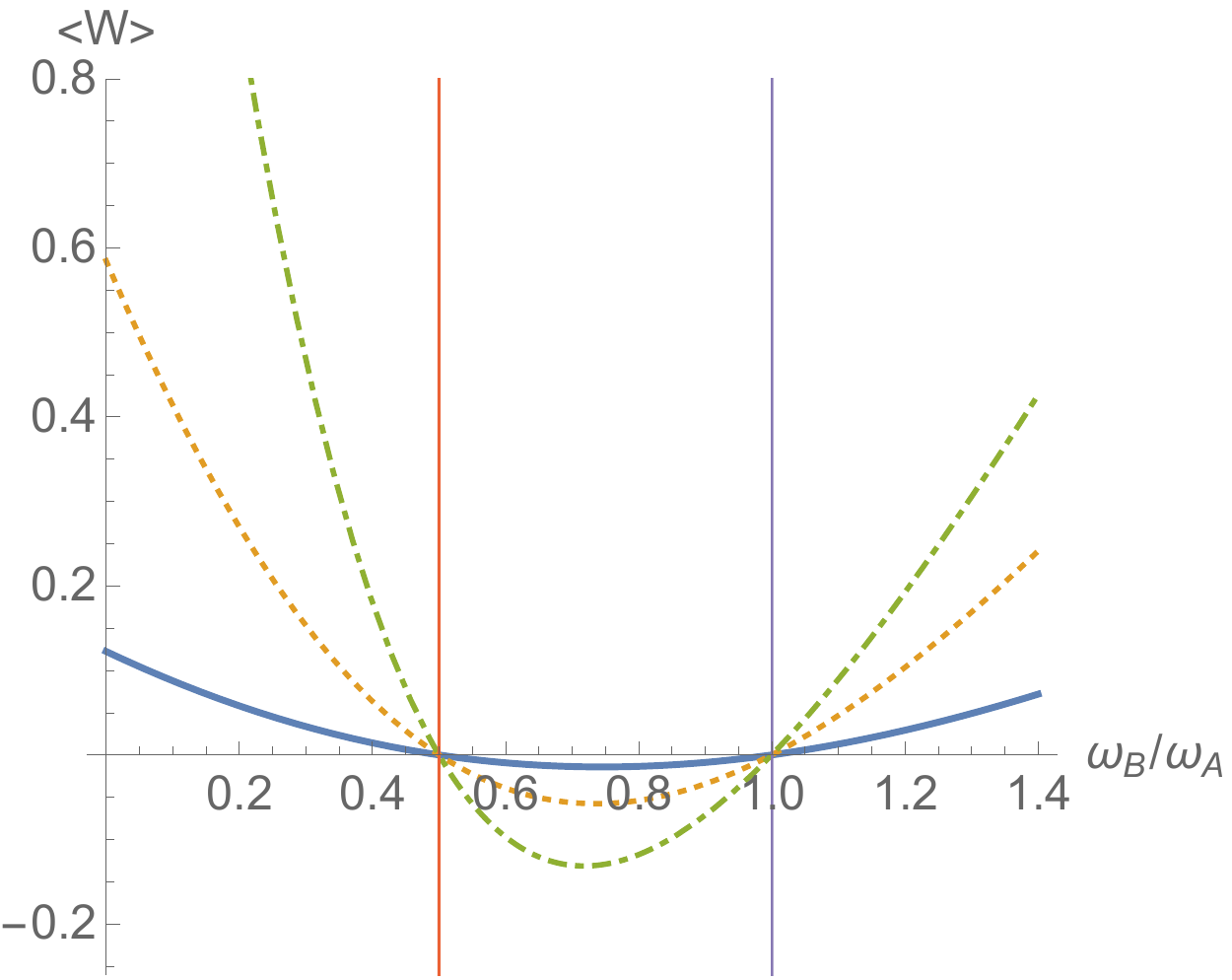}  
  \includegraphics[scale=0.45]{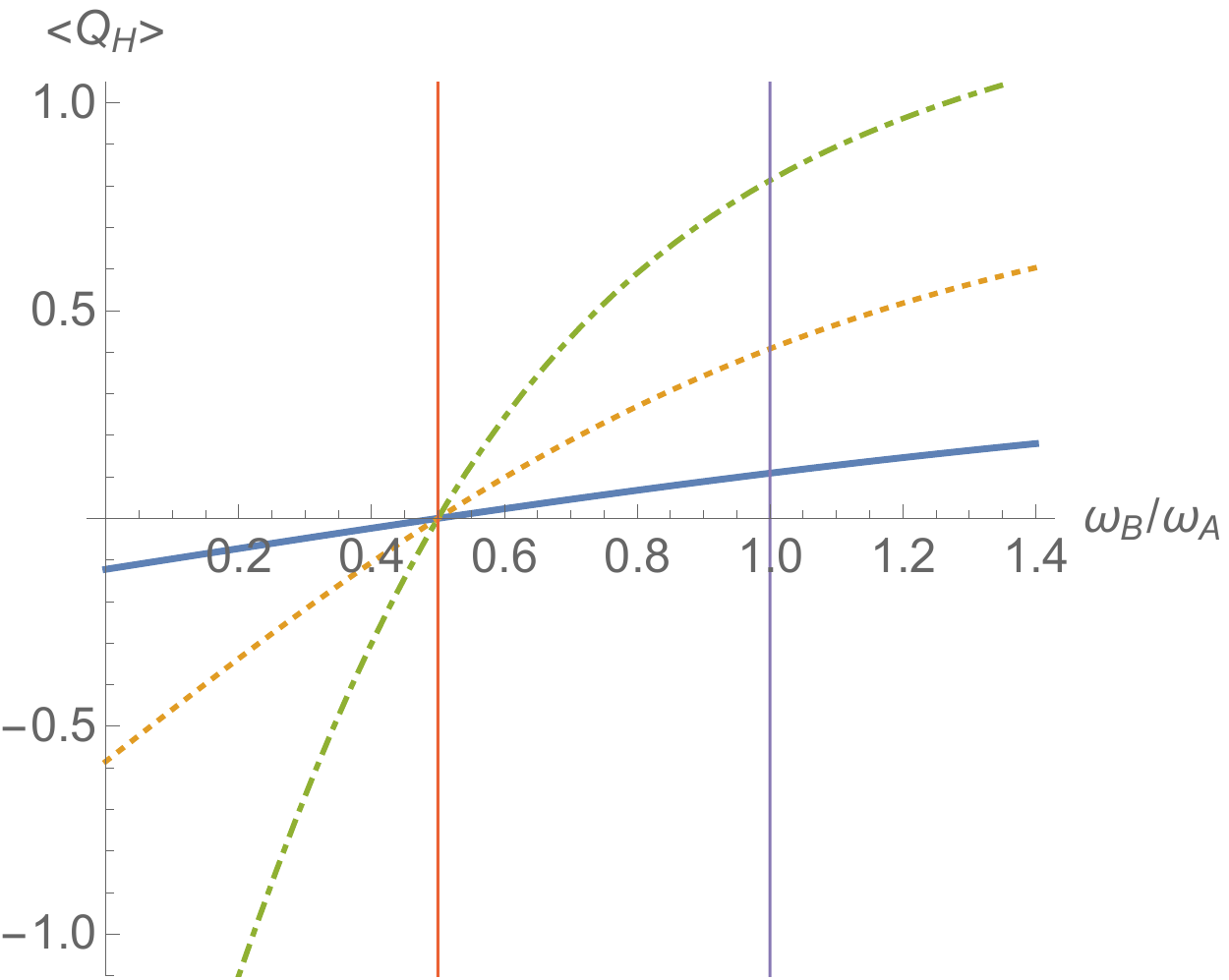}
  \includegraphics[scale=0.45]{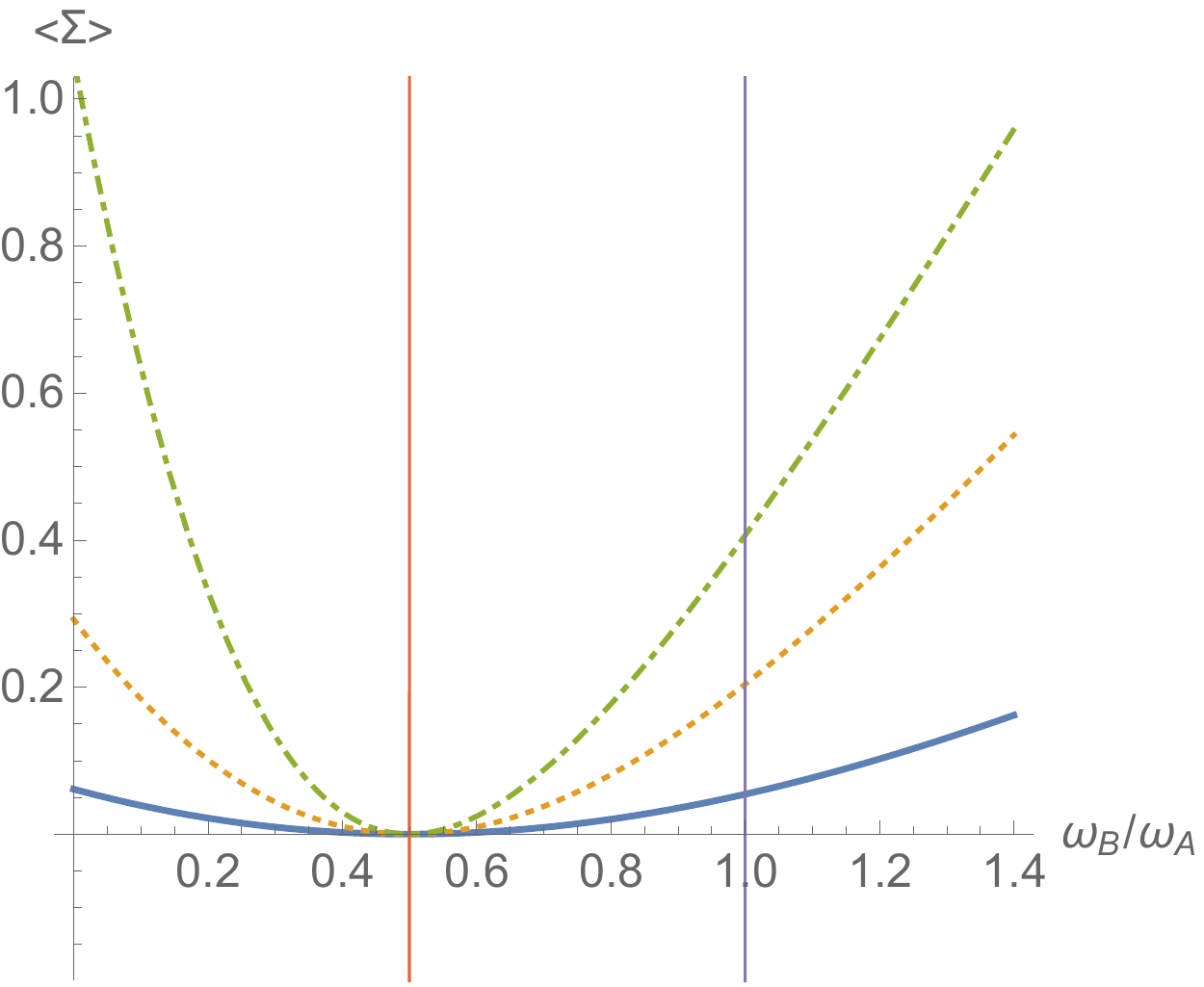}
\caption{Average work (left), heat (center) and entropy production
  (right) for parameters $T_A=2,\ T_B=1,\ \theta=
  \pi/2,\ \omega
_A=1$ versus $\omega _B/\omega _A$, for dimension $d=2$ (solid line),
  $d=4$ (dotted), and $d=8$ (dot-dashed). The three regimes of
  operation correspond to $\omega _B/\omega _A < \frac 12$
  (refrigerator), $\frac 12 < \omega _B/\omega _A <1 $ (heat engine), 
  and $\omega _B/\omega _A > 1 $ (thermal accelerator).}
\end{figure}
\par\noindent As expected, from Eq. (\ref{entp}) it follows that
the entropy production $\langle \Sigma \rangle$ is always positive.
Upon defining the mean occupation number
\begin{eqnarray}
N_X = \Tr \left[
  \frac{e^{-\beta _X H_X}}{Z_X} \frac{H_X}{\omega _X}\right ]
=\frac 12 \left [d-1 +
  \coth ( \beta _X \omega _X /2)  - d  \coth ( d\beta _X
  \omega _X /2)\right ] \equiv g(\beta _X \omega _X)
\;,\label{gx}
\end{eqnarray}
we can also rewrite concisely
\begin{eqnarray}
   \langle W \rangle =\sin^2\theta  (\omega _B -\omega _A) (N_A -N_B)
\;.\label{wco}
\end{eqnarray}
\par The efficiency $\eta $ of the heat engine is given by $\eta = \frac{\langle -
  W \rangle }{\langle Q_H \rangle } =1- \frac{\omega _B}{\omega _A}
\leq 1 - \frac {T_B}{T_A}\equiv \eta _C$, corresponding to the Otto
cycle efficiency. The Carnot efficiency $\eta _C$ is achieved only for
$\omega _A/\omega _B =T_A/T_B$ (i.e., with zero output
work).  Analogously, the coefficient of performance for the
refrigerator is given by $\zeta =\frac{\langle Q_C \rangle }{\langle W
  \rangle }=\frac{\omega _B}{\omega _A -\omega _B}\leq \frac {T_B}{T_A
  -T_B}=\zeta _C$. Notice that both $\eta $ and $\zeta $ are independent
of the coupling strength $\theta $, the temperature of the
reservoirs, and the dimension $d$.   
\par The present model shares some similarities with those studied in
Refs. \cite{strob,picc}, where the same Otto efficiency is achieved.
In particular, in Ref. \cite{picc} the interaction between
the working systems is modeled by a beamsplitter-like Hamiltonian, and
the same regimes of operation are obtained. In fact, in both models
the total excitation number of the systems is preserved by the working
interaction. Here, these feature is easily seen by the symmetry
relation
\begin{eqnarray}
  [H_I(t), H_A /\omega _A + H_B/\omega _B  ]=0
  \;.\label{feat} 
\end{eqnarray}
\par We can find (numerically) the efficiency $\eta _m$ at maximum work per
cycle. The result is plotted in Fig. 2, for different values of the
dimension $d$: one finds that $\eta _m$ is {\em larger} than
the Curzon-Ahlborn efficiency $\eta _{CA}=1-\sqrt{T_B/T_A}$, i.e. the efficiency
of the endoreversible Carnot cycle at maximum power \cite{ca1,ca4,ca2}
and rapidly converges to $\eta _{CA}$ for increasing values of the
dimension $d$ (see Appendix C for the limiting case $d \rightarrow
\infty$).  
\begin{figure}[ht]
  \includegraphics[scale=0.44]{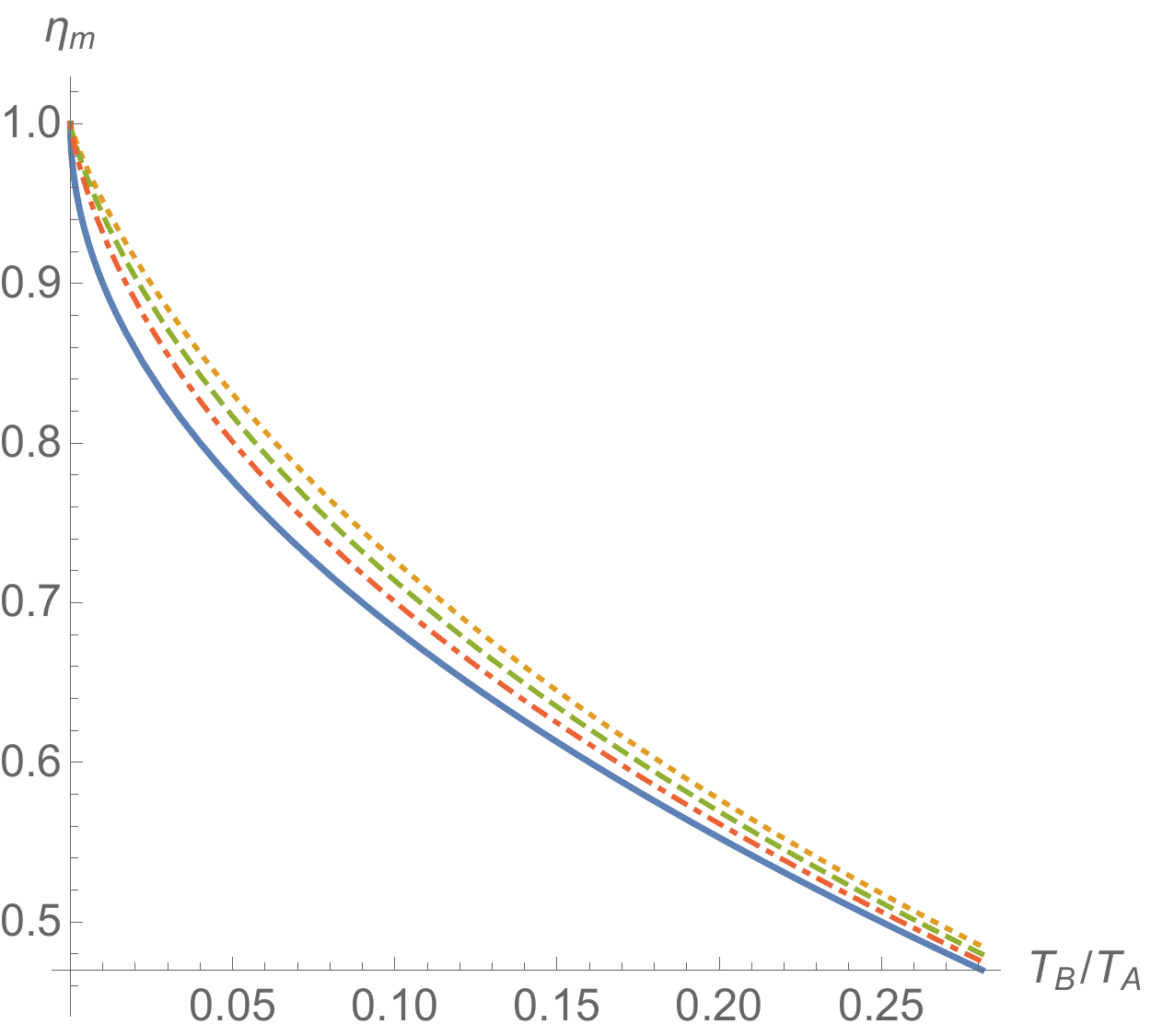}
\caption{Efficiency at maximum work $\eta _m$ versus the ratio
  $T_B/T_A$ for dimension $d=2$ (dotted), 4 (dashed), and 8
  (dot-dashed), along with the Curzon-Ahlborn curve $\eta
  _{CA}=1-\sqrt{T_B/T_A}$ in solid line.}
\end{figure}

\par By  the identity
$\frac{\beta _A \omega _A -\beta _B \omega _B}
  {\omega _A- \omega
    _B}= -\frac{1}{T_B}(\frac{\eta _C}{\eta }-1)$,
  one obtains the relation 
\begin{eqnarray}
   \langle \Sigma \rangle = \frac {\langle -W \rangle }{T_B} \left (\frac
           {\eta _C}{\eta }- 1 \right )\;\label{epp}
\end{eqnarray}
between average extracted work, entropy production and efficiency of
the heat engine. Analogously, one has $ \langle \Sigma \rangle = \frac
{\langle Q_C \rangle }{T_A} \left (\frac{1}{\zeta} -\frac{1}{\zeta
  _C}\right )$ for the refrigerator. Notice that these relations hold
also for the two-qubit and two-mode bosonic Otto engines
\cite{miobos}.
\section{Thermodynamic uncertainty relations and probability of
  stochastic work  and heat}
The second moments of work $\langle W^2 \rangle $, heat $\langle Q_H
^2 \rangle $, and the correlation $\langle W Q_H \rangle $ can be
obtained through Eqs. (\ref{mom}) and (\ref{chif}) by lengthy but
straightforward calculation, and one has
\begin{eqnarray}
    \langle W ^2\rangle
  &&= \frac {\sin ^2
  \theta}{2} (\omega _B -\omega _A)^2 
  \{d^2 -1 + \coth ^2(\beta _A \omega _A/2) + \coth ^2(\beta _B \omega _B/2)-   
  \coth (\beta _A \omega _A/2)   \coth (\beta _B \omega _B/2)
  \nonumber
\\& &
-d [\coth (\beta _A \omega _A/2) -\coth (\beta _B \omega _B/2) ][\coth
  (d\beta _A \omega _A/2) -
  \coth (d \beta _B \omega _B/2)] \nonumber \\& & - d^2 \coth (d
\beta _A \omega _A/2)\coth (d \beta _B \omega _B/2)
  \}
  \;,\label{w2}\\
  \langle Q_H ^2\rangle && = \frac {\omega _A ^2}{(\omega _B -\omega
    _A )^2} \langle W ^2\rangle \;,\\
  \langle W Q_H \rangle  && = \frac {\omega _A }{\omega _B -\omega
    _A }  \langle W ^2\rangle \;.
\end{eqnarray}
From the above equations, along with Eq. (\ref{entp}),  we find
an exact identity relating the inverse signal-to noise ratios
and the entropy production, namely
\begin{eqnarray}
  \frac{\mbox{var}(W)}{\langle W \rangle ^2}
  =  \frac{\mbox{var}(Q)}{\langle Q_H \rangle ^2}
=  \frac{\mbox{cov}(W,Q_H)}{\langle W \rangle \langle Q_H\rangle }
=\frac {(\beta _B \omega _B -\beta _A \omega _A )
  f(\beta _A \omega _A,\beta _B \omega _B  ,d)}{\langle \Sigma \rangle } -1
\;,\label{invw}
\end{eqnarray}
with
\begin{eqnarray}
f(x,y,d)&&=
\{d^2 -1 + \coth ^2(x/2) + \coth ^2(y/2)-   
  \coth (x/2)   \coth (y/2)
  \nonumber
\\& &
-d [\coth (x/2) -\coth (y/2) ][\coth
  (d x/2) -
  \coth (d y /2)] - d^2 \coth (d
x/2)\coth (d y/2) \}\nonumber \\& &
\times 
 \{ \coth ( x/2) -    \coth ( y /2) - d [\coth ( d x /2) -\coth ( d y /2)]    
   \}^{-1}
\;.\label{fd}
\end{eqnarray}
We note that the only dependence on the coupling parameter $\theta $
comes for the inverse of the average entropy production $\langle
\Sigma \rangle$. Hence, the above ratios are minimized versus $\theta
$ for $\theta =\frac \pi 2$, for which also the entropy production
achieves the maximum. Then, the reduction of the noise-to-signal ratio
associated to work extraction (or cooling performance) comes at a
price of increased entropy production.  We observe that for $\theta
=\pi/2$ the unitary stroke transforms the initial bi-thermal state
(\ref{eqq}) with inverse temperatures $\beta _A$ and $\beta _B$ into a
bi-thermal state with final inverse temperatures $ \beta _B \omega
_B/\omega _A$ and $\beta _A \omega _A/ \omega _B$ without leaving
final correlations between the two qudits. On the other hand,
operating at zero entropy production (i.e., for $\beta _A \omega _A
\rightarrow \beta _B \omega _B$, thus approaching the Carnot
efficiency) will induce a divergence in Eq. (\ref{invw}). 
\par For the heat-engine regime $\beta _A \omega _A < \beta _B \omega _B$
and fixed values of $d$ and $\theta $,
numerical inspection shows that for assigned value of $\beta _A \omega
_A$  (of $\beta _B \omega _B$) the ratio
$\frac{\mbox{var}(W)}{\langle W \rangle ^2}$ is  minimized for $\beta
_B \omega _B \rightarrow \infty $
(for $\beta _A \omega _A
\rightarrow0$), and the ultimate minimization is 
given by
\begin{eqnarray}
\frac{\mbox{var}(W)}{\langle W \rangle ^2}= \frac{d+1 +3
  (d-1)\cos ^2 \theta}{3(d-1) \sin ^2 \theta}
\;,
\end{eqnarray}
achieved for $\beta _A \omega _A \rightarrow0$ and $\beta_B \omega _B
\rightarrow \infty $.

\par Note that for both the heat engine and the refrigerator the sign
of $\mbox{cov}(W,Q_H)$ is negative, whereas for the thermal
accelerator where external work is supplied to increase the heat flow
from hot to cold reservoir the covariance is positive.

\par For $d=2$, since the function $g$ in Eq. (\ref{gx}) is 
easily inverted, namely $\beta _X \omega _X = \ln \frac{1-N_X}{N_X}$,
Eq. (\ref{w2}) can be rewritten as
\begin{eqnarray}
\langle W^2 \rangle = \sin ^2 \theta (\omega _B -\omega _A)^2 (N_A +N_B -2 N_A N_B)\;,
\end{eqnarray}
thus recovering the result for qubits
\cite{miobos}. Also Eq. (\ref{fd}) simplifies as 
\begin{eqnarray}
f(x,y,2)=\coth[(y-x)/2]\;.
\end{eqnarray}
From numerical evidence one has $(y-x)f(x,y,d)\geq 2$, with equality
in the limit $x\rightarrow y$, and hence the following TUR is
obtained 
\begin{eqnarray} 
\frac{\mbox{var}(W)}{\langle W \rangle ^2} \geq \frac {2}{\langle \Sigma \rangle}-1  
\;,\label{bb2}
\end{eqnarray}    
which holds for all parameters and any dimension $d$.  
As in the two-qubit case, the presence of the $-1$ term implies that 
the standard TUR $\frac{\mbox{var}(W)}{\langle W \rangle ^2} \geq
\frac {2}{\langle \Sigma \rangle}$ can be slightly violated
\cite{timpa,miobos}. 
Remarkably, a similar small violation has been recently reported in
Ref. \cite{prec} for a different model of thermal machine (i.e. a
two-qubit steady-state and autonomous engine), where also a stronger
violation is found for a three-qubit model. 

\par In Fig. 3 we report the signal-to-noise ratio $ \langle W \rangle
^2 /\mbox{var}(W)$ along with the function $\langle \Sigma \rangle /2$
for the cases $d=2$ and $d=8$ with $\theta =\pi/2$. We generally
observe that the region of parameter space for a violation of the
standard TUR is shrunk for increasing values of the dimension
$d$. This is more apparent in Fig. 4, where the product
${\mbox{var}(W)}\langle \Sigma \rangle /\langle W \rangle ^2$ is
reported versus $\beta _B \omega _B$ for fixed $\beta _A \omega
_A=0.1$, coupling $\theta =\pi/2$ and dimension $d=2$, 4, and
6. Indeed, we recall that for the two-mode bosonic engine the standard
TUR is never violated, since $\frac{\mbox{var}(W)}{\langle W \rangle
  ^2} \geq \frac {2}{\langle \Sigma \rangle}+1$ holds \cite{miobos}.
A similar shrinking effect is observed also for decreasing values of
the coupling $\theta $ at fixed dimension $d$, along with a rapid
decrease of the strength of the violations, as shown for example in
Fig. 5 for the case $d=2$ with coupling values $\theta =\pi/2$, $5\pi
/12$, and $\pi /3$. We report that the strongest violation of the
standard TUR is numerically obtained for $d=2$, $\theta =\pi/2$,
$\beta _A \omega _A \rightarrow 0$, and $\beta _B \omega _B \simeq
2.010$, for which ${\mbox{var}(W)}\langle \Sigma \rangle /\langle W
\rangle ^2 \simeq 1.864$.  

\begin{figure}[ht]
\begin{subfigure}
  \centering
  \includegraphics[scale=0.52]{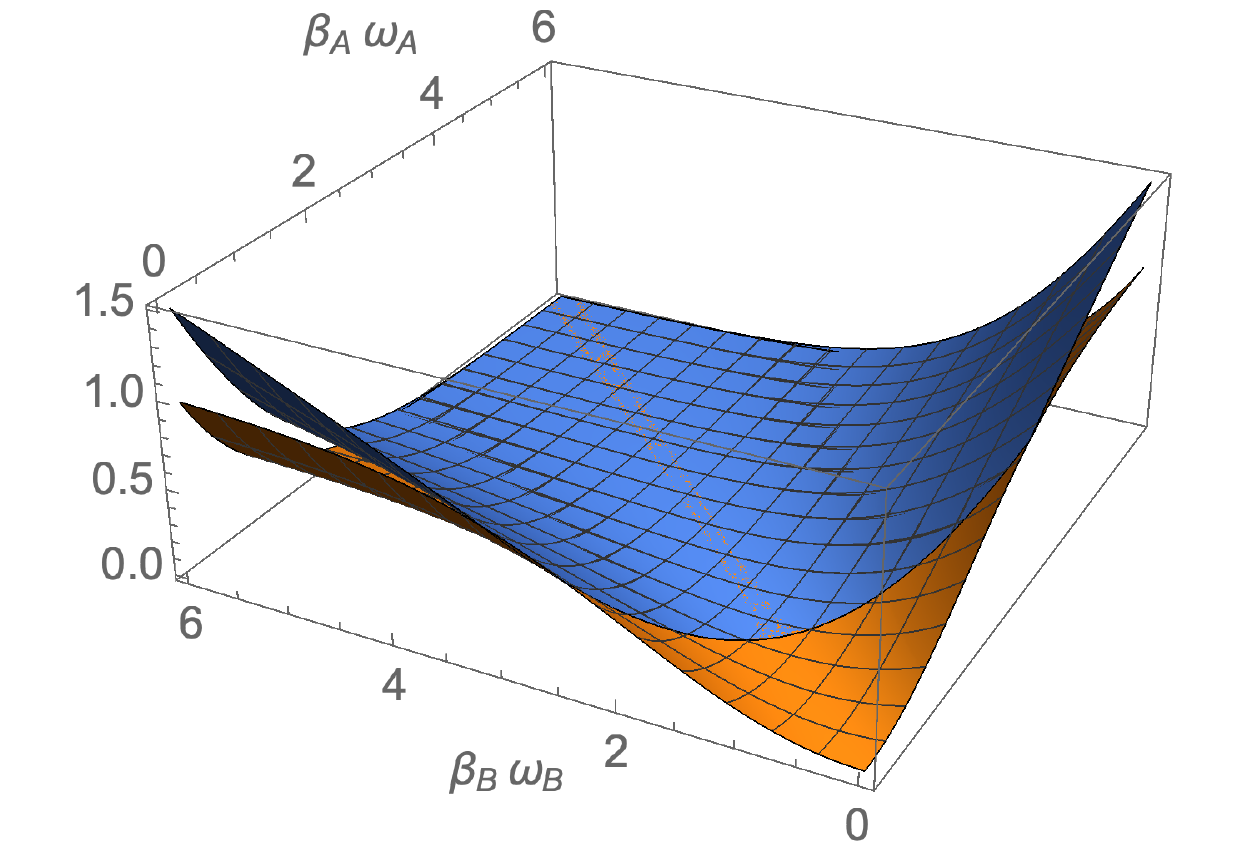}
\end{subfigure}
\ \ \ \ \ \ \ \ \ 
\begin{subfigure}
  \centering
  \includegraphics[scale=0.52]{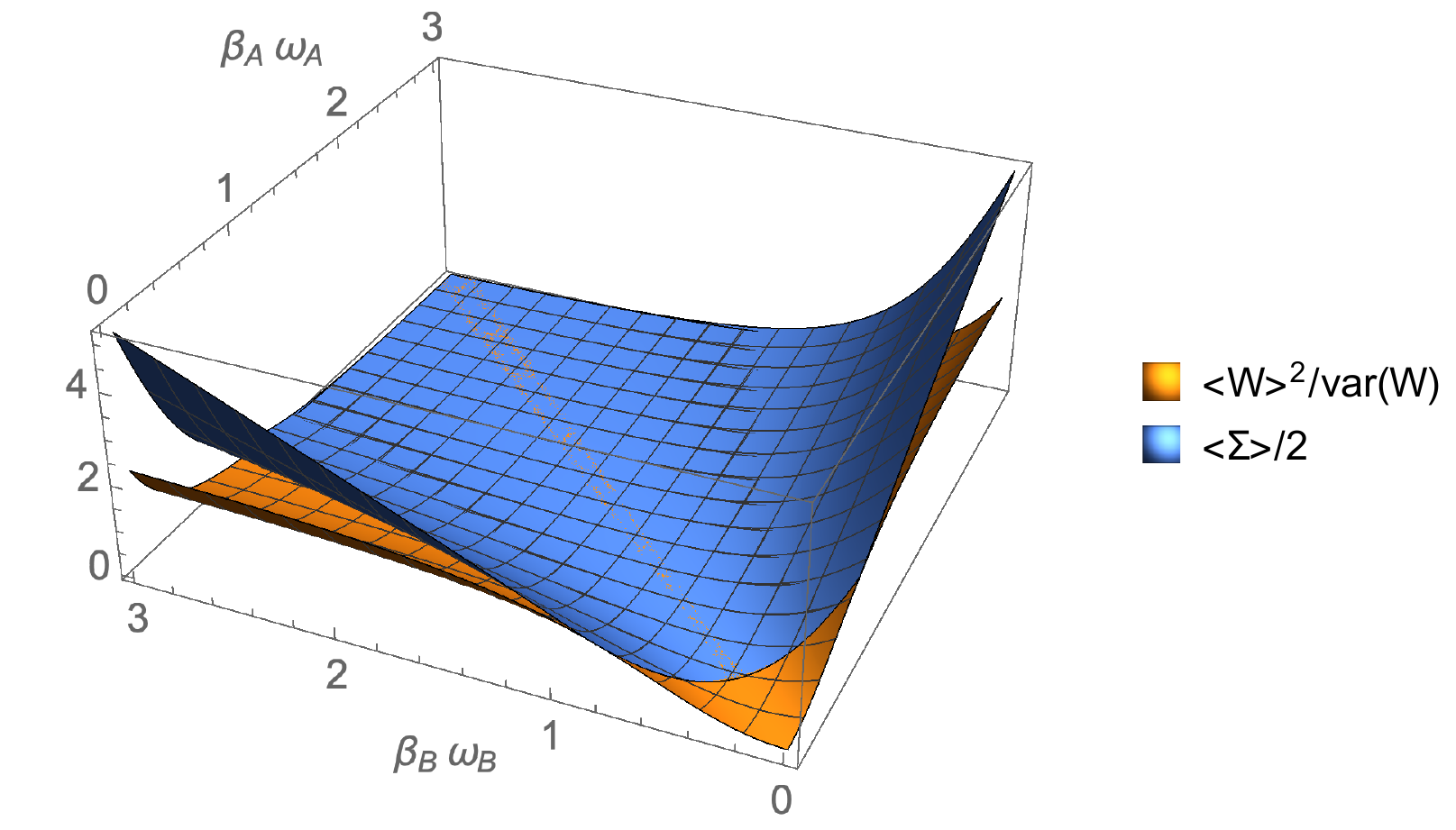}
\end{subfigure}
\caption{Plot of the signal-to-noise ratio of work
  $  \langle W
  \rangle ^2 /\mbox{var}(W)$ and scaled entropy production $\langle \Sigma
    \rangle /2$ with coupling $\theta =\pi/2$  for dimension $d=2$ (left) and
    $d=8$ (right) as a function of parameters $\beta _A
    \omega _A$ and
    $\beta _B \omega _B$. The region of parameter space where the 
    standard TUR is violated is shrunk for increasing dimension of the working
    systems.} 
\end{figure}
\begin{figure}[ht]
  \includegraphics[scale=0.49]{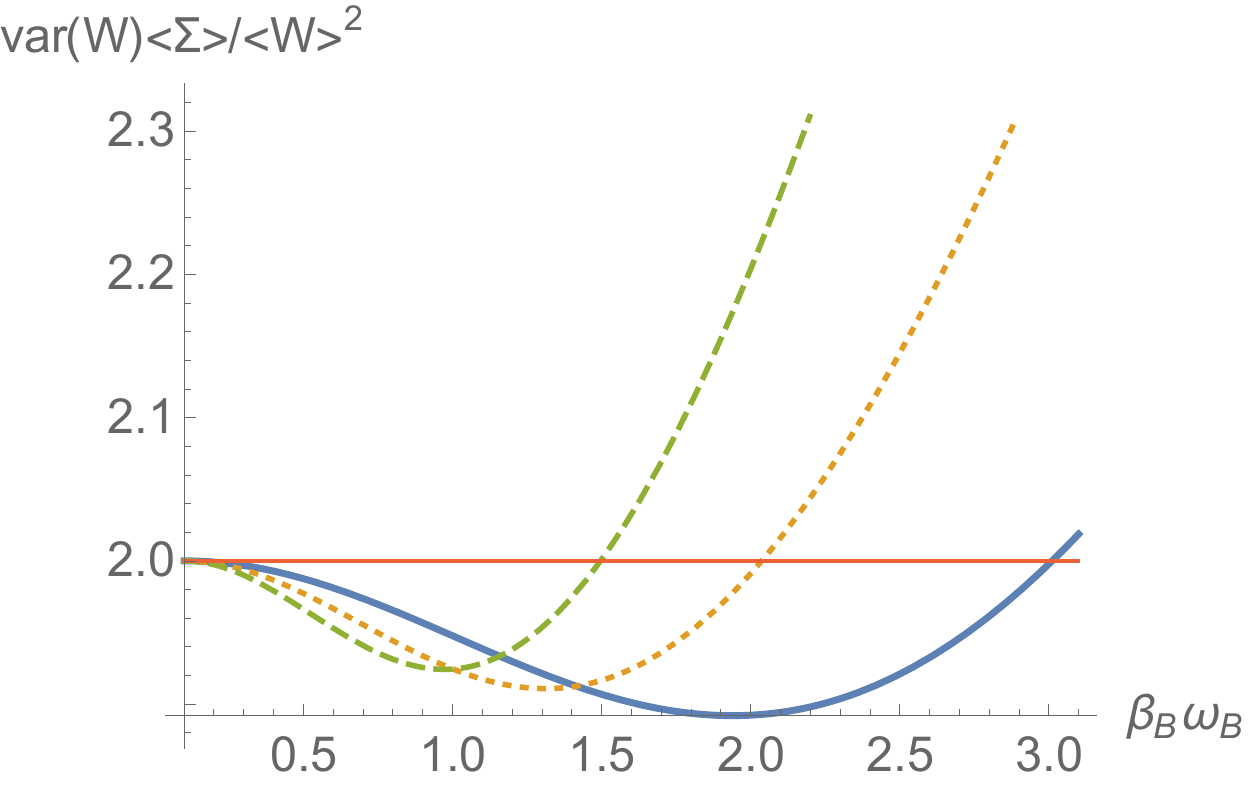}
  \caption{Ratio ${\mbox{var}(W)}\langle \Sigma \rangle /\langle W
      \rangle ^2$ versus $\beta _B \omega _B$ with $\beta _A \omega
  _A=0.1$, $\theta =\pi/2$, and dimension $d=2$ (solid line), 3
      (dotted), and 4 (dashed).} 
\end{figure}
\begin{figure}[h!]
  \includegraphics[scale=0.49]{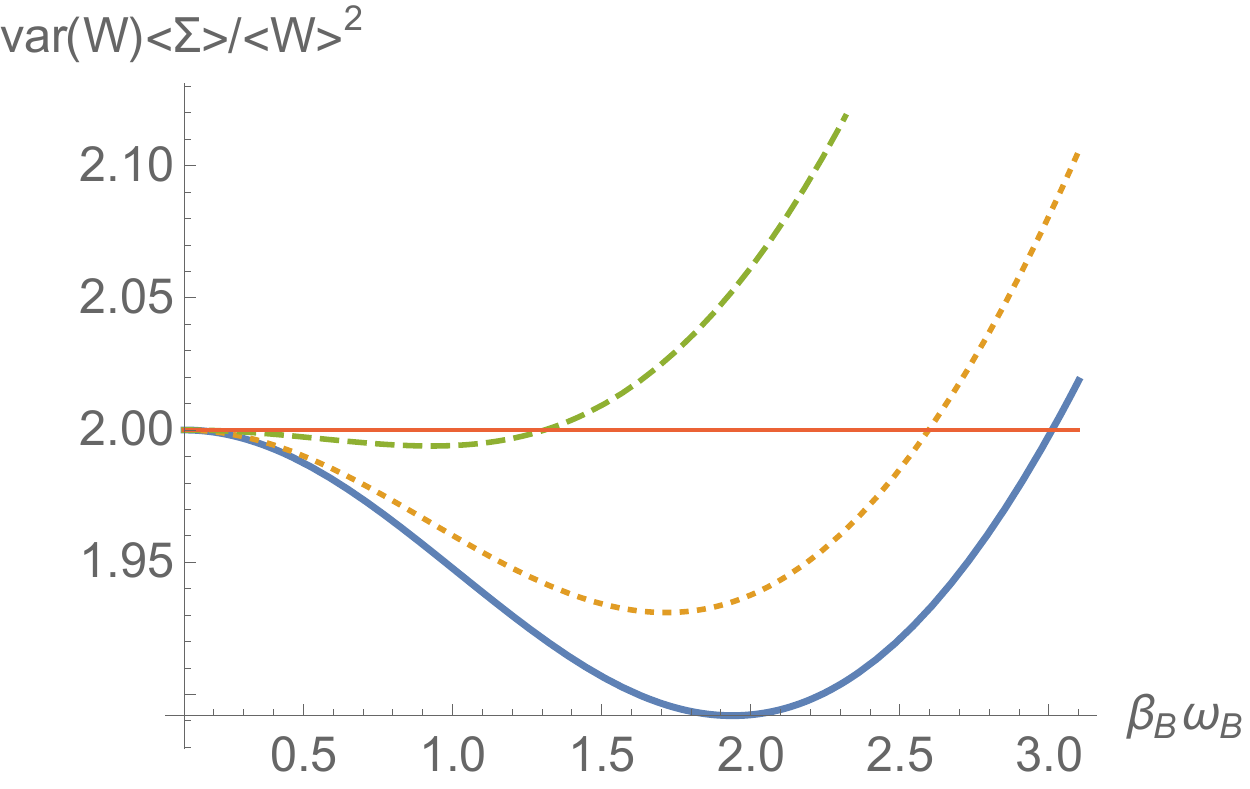}
  \caption{Ratio ${\mbox{var}(W)}\langle \Sigma \rangle /\langle W
      \rangle ^2$ versus $\beta _B \omega _B$ with $\beta _A \omega
  _A=0.1$, $d=2$, and coupling strength $\theta =\pi/
      2$ (solid line), $5\pi/12$ (dotted), and 
        $\pi /3$ (dashed).} 
\end{figure}
\par From Eq. (\ref{epp}) and the bound $\frac{\langle W^2 \rangle
}{\langle W \rangle ^2} \geq \frac {2}{\langle \Sigma \rangle}$
equivalent to Eq. (\ref{bb2}), we can obtain the following relation
between the average extracted work, second moment and efficiency
\begin{eqnarray}
\langle -W \rangle \leq \frac {\langle W^2 \rangle }{2 T_B} \left (\frac {\eta
  _C} {\eta} -1 \right )  
\;,\label{eqpo}
\end{eqnarray}
which can also be written as a bound on the efficiency,
namely
\begin{eqnarray}
\eta \leq \frac{\eta _C}{1+ 2T_B \langle -W \rangle / \langle W^2
  \rangle }\;. \label{eqpo2}
\end{eqnarray}
Hence, in order to increase the efficiency, one must either reduce the
output work or increase the second moment of work distribution, thus
undermining the engine reliability. For the two-mode
bosonic Otto engine analogous equations as (\ref{eqpo}) and
(\ref{eqpo2}) hold \cite{miobos}, when replacing $\langle W^2 \rangle
$ with $\mbox{var}(W)$.
\par Since the characteristic function is periodic in $\lambda $ and
$\mu $ with period $\frac{2\pi}{|\omega _A -\omega _B|}$ and
$\frac{2\pi}{ \omega _A}$ the joint probability 
$p(W,Q_H)$ of the stochastic work and heat is discrete, with
$W$ and $Q_H$ as integer multiples of $\omega _A -\omega _B$ and $ \omega
_A$, respectively. Then one has
\begin{eqnarray}
  p[W=m(\omega _A -\omega _B), Q_H =  n \omega _A]= \frac {
    \omega _A |\omega
    _A - \omega _B|}{(2 \pi)^2} 
\int _{-\frac{\pi}{|\omega _A -\omega _B |}}^
{\frac{\pi}{|\omega _A -\omega _B|}} d \lambda
\int _{-\frac{\pi}{ \omega _A }}^
     {\frac{\pi}{ \omega _A}} d \mu \, \chi (\lambda ,\mu ) 
     e^{-i m (\omega _A -\omega _B)\lambda  -i  n \omega _A \mu }
\;.\label{ptf} 
\end{eqnarray}  
Since $\chi(\lambda , \mu)$ is a function of the single
variable $ \xi = (\omega _A -\omega _B) \lambda - \omega _A \mu $,
namely $\chi (\lambda ,\mu)=\chi \left (0,\mu - \frac {(\omega _A -\omega
  _B)\lambda }{\omega _A}\right )$, by the 
change of variables $\mu \rightarrow \omega _A \mu -(\omega _A -\omega
_B) \lambda$ and $\lambda \rightarrow (\omega _A -\omega _B) \lambda $
in Eq. (\ref{ptf}) we obtain
\begin{eqnarray}
&&  p[W=m(\omega _A -\omega _B), Q_H =  n \omega _A]=
  \frac {1}
{(2 \pi)^2} 
\int _ {-\pi}^\pi 
     d \lambda
\int _{-\pi -\lambda }
     ^{\pi -\lambda } d \mu \,   \chi \left (0,
 \frac{\mu}{ \omega
   _A} \right )\,
     e^{-i (n+m) \lambda  -i  n  \mu }\nonumber \\& & =
  \int _{0}^{2\pi} \frac{d \mu}{2\pi}
\,  \chi \left (0,
 \frac{\mu}{ \omega
    _A} \right ) e^{-i
              n \mu  } 
  \int _{-\pi }^{\pi} \frac{d \lambda }{2\pi}
  \,   e^{-i (n+m) \lambda  }=
   \delta _{m,-n} \, 
  \int _{0}^{2\pi} \frac{d \mu}{2\pi}
\,  \chi \left (0,
 \frac{\mu}{ \omega
   _A} \right )
 e^{-i
              n \mu  } 
     \;.  \label{19}
\end{eqnarray}
This means that the stochastic work and heat are perfectly correlated,
i.e. 
\begin{eqnarray}
  &&  p[W=m(\omega _A - \omega _B), Q_H =  n \omega _A]=p[W=m(\omega
    _A - \omega _B)] \delta _{n,-m}= 
  p[Q_H=  n \omega _A]  \delta _{m,-n}\;.\label{sy}
\end{eqnarray}
This feature is due to the symmetry (\ref{feat})
and implies that $\langle (-W/Q_H)^n \rangle =\langle -W/Q_H \rangle ^n =\left ( 1-
\frac {\omega _B }{\omega _A}\right )^n $, namely there are no
efficiency fluctuations.
\par The simplest way to find explicitly the probability is to proceed
from Eq. (\ref{19}) by using the expression of the
characteristic function in the last line of Eq. (\ref{chif9}), and one
obtains 
\begin{eqnarray}
  p[Q_H =n \omega _A]&&=
\int _{0}^{2\pi} \frac{d \mu}{2\pi}\,
\left (\cos^2 \theta  e^{-i n \mu }
 + \sin^2
 \theta \frac{1}{Z_A Z_B} \sum_{l,s=0}^{d-1} e^{-l \beta _A \omega _A}
 e^{-s \beta _B \omega _B}  e^{i (l-s-n)\mu } \right )
\nonumber \\& &= \delta _{n,0} \cos^2 \theta  + 
  \sin^2
  \theta \frac{1}{Z_A Z_B} \sum _{s=\max \{0,-n \}}^{d-1-\max \{0,n\}}
 e^{-s (\beta _A \omega _A + \beta _B \omega _B)} e^{-\beta _A \omega _A n}
\;,
\end{eqnarray}
which can be summarized as
\begin{eqnarray}
   p[Q_H =n \omega _A]
=\delta _{n,0} \cos^2 \theta + 
  \sin^2
 \theta \frac{1}{Z_A Z_B} \frac{1- e^{-(d-|n|)(\beta _A \omega _A+
     \beta _B \omega _B)}}
   {1- e^{-(\beta _A \omega _A+
     \beta _B \omega _B)}}\times 
\left\{
\begin{array}{ll}
e^{-\beta _A \omega _A n} & \qquad \mbox{for }0 \leq n \leq d-1 
  \;,\\
  e^{-\beta _B \omega _B |n|} &
     \qquad
\mbox{for }1-d \leq n < 0\;. 
\end{array}
\right.
   \;\label{clos}
\end{eqnarray}
\par In Fig. 6 we report the probability  for the stochastic work in
$(\omega _A - \omega _B)$ units for $\beta _A \omega _A =1$ and $\beta
_B \omega _B =2$ and dimension $d=8$,    
pertaining to three increasing values of the strength interaction, i.e. $\theta
=\pi/4,\ \pi/3,$ and  $\pi/2$.  
\begin{figure}[ht]
  \includegraphics[scale=0.46]{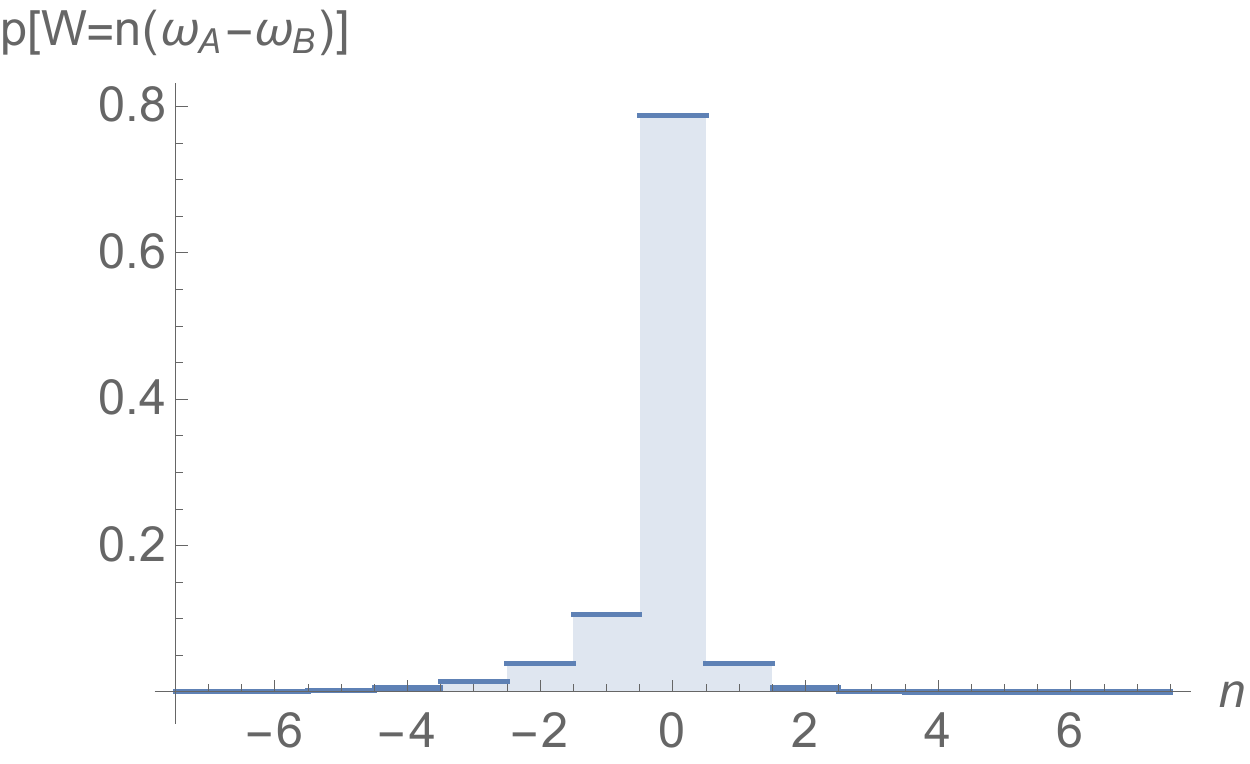}
  \includegraphics[scale=0.46]{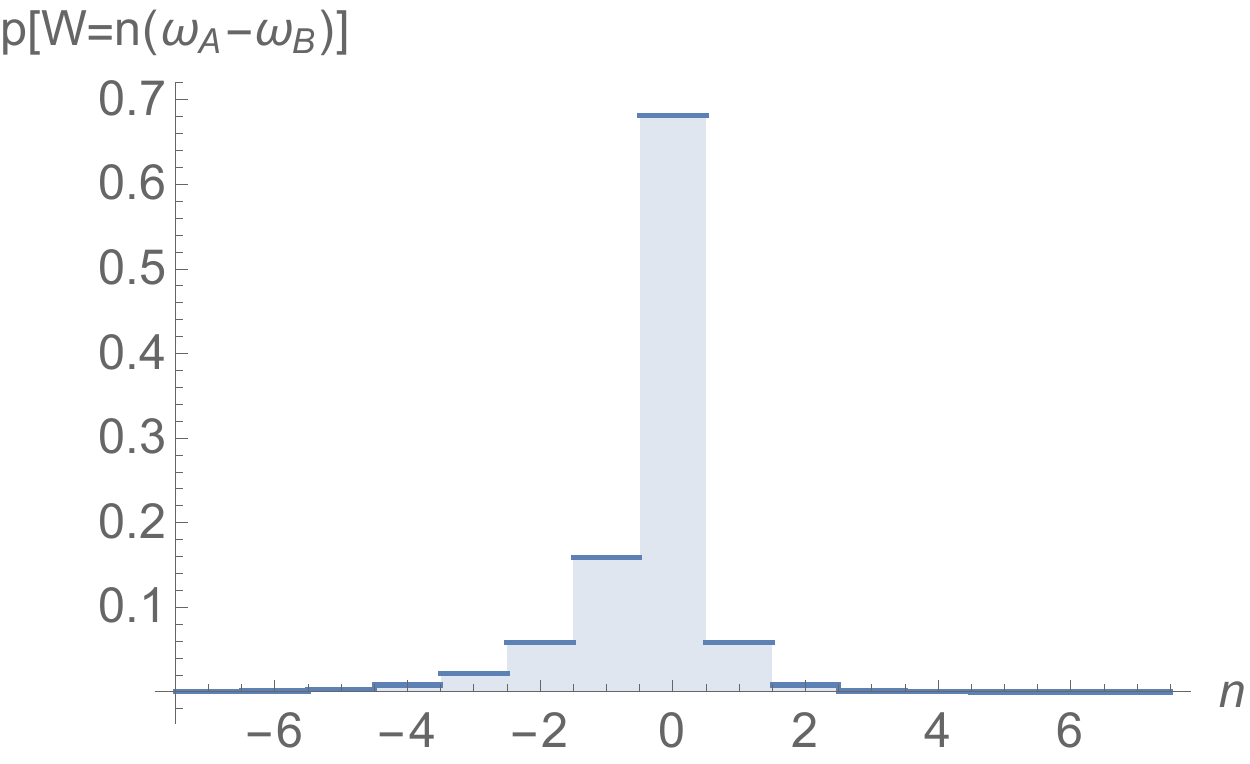}
  \includegraphics[scale=0.46]{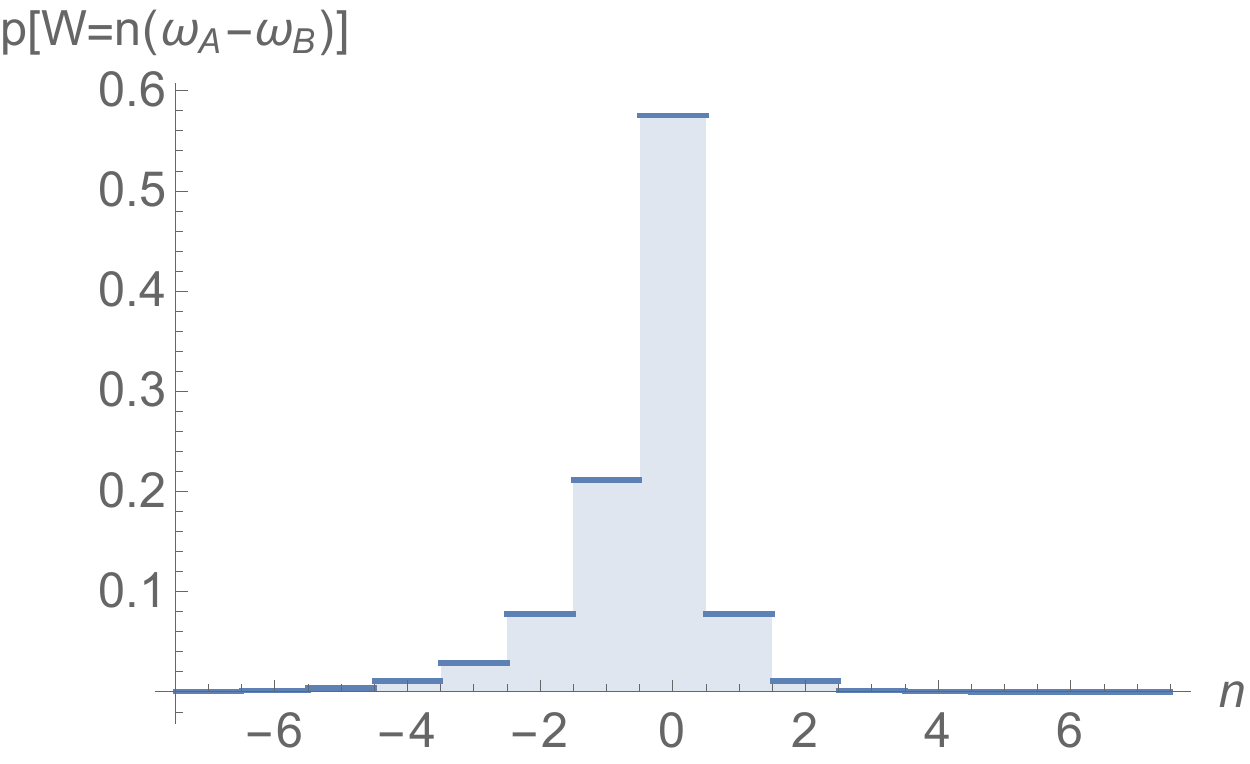}
\caption{Probability of the stochastic work in $\omega _A -\omega _B$ units, for
  $\beta _A \omega _A =1$ and $\beta _B \omega _B =2$ and dimension
  $d=8$, with strength interaction $\theta =\pi/4$ (left), $\pi/3$
  (center), $\pi/2$ (right).  By exchanging $n \rightarrow -n$, the
  same histograms represent the distribution of heat released by the
  hotter reservoir in $\omega _A $ units [see Eq. (\ref{sy})].}
\end{figure}
\par The closed form for the probability in Eq. (\ref{clos}) allows one
to explicitly verify the detailed fluctuation theorem in  Eq. (\ref{dett})
as follows
\begin{eqnarray}
  \frac{p[W=-n(\omega _A -\omega _B),Q_H =n\omega _A]}
    {p[W=n(\omega _A -\omega _B),Q_H =-n\omega _A]}=
       e^{(\beta_B\omega _B -\beta _A \omega _A)n}
    =
  e^{(\beta _B -\beta _A)n \omega _A - \beta _B n (\omega _A -\omega
    _B)}=
e^{(\beta _B- \beta _A )Q_H +\beta _B W}\;.
\end{eqnarray}
\section{Finite-time analysis for perfect-swap strokes}
The thermal strokes considered in the previous sections implicitly
assume infinite duration in order to guarantee a complete relaxation
of the qudits by weak coupling to the temperatures of the respective
thermal reservoirs. This means that indeed the output power per cycle
becomes vanishing. However, the efficiency for a working cycle at
finite times and hence with non-zero output power is usually of great
practical importance \cite{nano3,otto4,ca1,ca4,ca2,ca5,ca3,ca6}.  \par In
this section we provide a preliminary study of the finite-time
performance of the presented model for the specific case of
perfect-swap operation $\theta =\pi/2$, postponing a general analysis
for future work. In fact, as previously noticed after Eq. (\ref{fd}),
a bi-thermal state of the qudits remains factorized and bi-thermal
under perfect-swap operation.  Then, we can provide a simple model for
the effect of partial thermalization in the case of finite-time
stroke, without resorting to a specific master-equation
approach. After a number of transient cycles, the state of the two
qudits will rapidly achieve the steady state of the map given by the
composition of the swap stroke followed by the thermal stroke. Indeed,
let us consider a time-dependent relaxation of the mean occupation
number for both qudits towards their respective equilibrium values
$N_A$ and $N_B$ as
\begin{eqnarray}
\frac{d N_X (t)}{dt}= -\alpha _X [N_X(t) -N_X]\;
\end{eqnarray}
for $X=A,B$, where ${\alpha _X}$ denotes the relaxation rate constants
and $N_X$ are given by Eq. (\ref{gx}).  Clearly, we have $N_X(t)=[N_X
  (0)-N_X]e^{-\alpha _X t}+N_X$.  By taking thermal strokes with
finite duration $\tau _q$, at the end of the $(n + 1)$-th cycle the
state will be bi-Gibbsian and the mean occupation numbers will satisfy
the recursive relations
\begin{eqnarray}
&&N_A^{(n+1)}=(N_B ^{(n)}-N_A)e^{-\alpha _A
    \tau _q}+N_A\;,\nonumber \\ 
&&N_B^{(n+1)}=(N_A ^{(n)}-N_B)e^{-\alpha _B
  \tau _q}+N_B\;.  
\end{eqnarray}
The cycles lead to a periodic state corresponding to the steady-state
solution given by
\begin{eqnarray}
&&N_A ^*=\frac{N_A (1-e^{- \alpha _A \tau _q}) +N_B (1-e^{-
      \alpha _B \tau _q})
    e^{- \alpha _A \tau _q}}{1-e^{-( \alpha _A
    + \alpha _B)\tau _q}}
    \;,\nonumber \\
&&N_B ^*=\frac{N_B (1-e^{- \alpha _B \tau _q}) +N_A
      (1-e^{- \alpha _A \tau _q}) e^{- \alpha _B
        \tau _q}}{1-e^{-( \alpha _A
    + \alpha _B)\tau _q}}
  \;.\label{sn}    
\end{eqnarray}
From Eqs. (\ref{sn}) we also note the symmetry relation 
\begin{eqnarray}
N_A^* -N_B ^*
  =\frac{(1-e^{- \alpha
    _A \tau _q})(1-e^{- \alpha _B \tau _q})}{1-e^{-( \alpha _A
      + \alpha
    _B)\tau _q}}  \left  ( N_A -N_B  \right )
\;.\label{dif}
\end{eqnarray}
For finite-time thermalization strokes, after a transient, the
characteristic function in the limit cycle is then still given by
Eq. (10) by replacing $\beta _X$ with the effective inverse
temperatures $\beta_X^* =g^{-1}(N_X^*)/\omega _X$ in terms of the
inverse of the function $g$ given in Eq. (\ref{gx}). The average work
in the steady cycles is then given by
\begin{eqnarray}
\langle W \rangle =(\omega _B -\omega _A)(N_A^* -N_B^*)\;,
\end{eqnarray}
and for Eq. (\ref{dif}) a simple scaling factor appears with respect
to Eq. (\ref{wco}). Since the
entropy production is ascribed to the temperature of the thermal baths
one has
\begin{eqnarray}
  \langle \Sigma \rangle =(\beta _B \omega _B -\beta _A \omega
  _A)(N^*_A - N^*_B)
  \;,
\end{eqnarray}
and so Eq. (\ref{entp}) still holds. The second moment in
Eq. (\ref{w2}) and the joint probability of the stochastic work and
heat in Eq. (\ref{clos}) are obtained by the replacement $(\beta _A ,
\beta _B ) \rightarrow (\beta _A^* , \beta _B^* )$. Hence, the
efficiency of the swap engine remains a non-fluctuating quantity even
in the finite-time regime and is still given by $\eta = 1-
\frac{\omega _B}{\omega _A}$.

By neglecting the unitary stroke duration (e.g., we can take $\kappa \gg 1$
and $\tau _w \ll 1$, with $\theta =\kappa \tau _w$ as an odd multiple
of $\pi/2 $), the output power
per cycle is given by
\begin{eqnarray}
P=\langle - W \rangle /\tau _q=   
(\omega _A -\omega _B)(N^*_A -N^*_B)/\tau _q 
\;.
\end{eqnarray}
By assuming  for simplicity equal relaxation rates $ \alpha
_A =  \alpha _B=  \alpha $ for the two reservoirs, one has
\begin{eqnarray}
P=\frac {\mbox{tanh}(\alpha \tau _q/2)}{\tau _q}(\omega _A -\omega _B)(N_A -N_B)
  \;,\label{pfat}
\end{eqnarray}
which is trivially maximized versus $\tau _q$ for $\tau _q\rightarrow
0$, giving finite power $P=\frac \alpha 2 (\omega _A -\omega _B)(N_A
-N_B)$.  This means that the maximum power is achieved by a bang-bang approach
with very short-term strokes.  By dropping the condition $\tau _w \ll
1$, notice that the optimal power will be smaller and obtained for a
non-zero finite value of $\tau _q$, as shown in Fig. 7. These
considerations are analogous with the results of Ref. \cite{otto4}
pertaining to finite-time four-stroke Otto engines. Notice that the
previous results about the efficiency at maximum work (see Fig. 2)
apply at maximum output power as well, since the dependence on
time in Eq. (\ref{pfat}) is simply given by a factor and the
efficiency is unchanged and independent of the cycle duration. We note
that the possibility of surpassing the Curzon-Ahlborn efficiency has
been recognized in a number of different scenarios
\cite{nano3,cpf,ca5,ca3,ca6,erd,picc,hofer,bera}.  
\begin{figure}[ht]
  \includegraphics[scale=0.52]{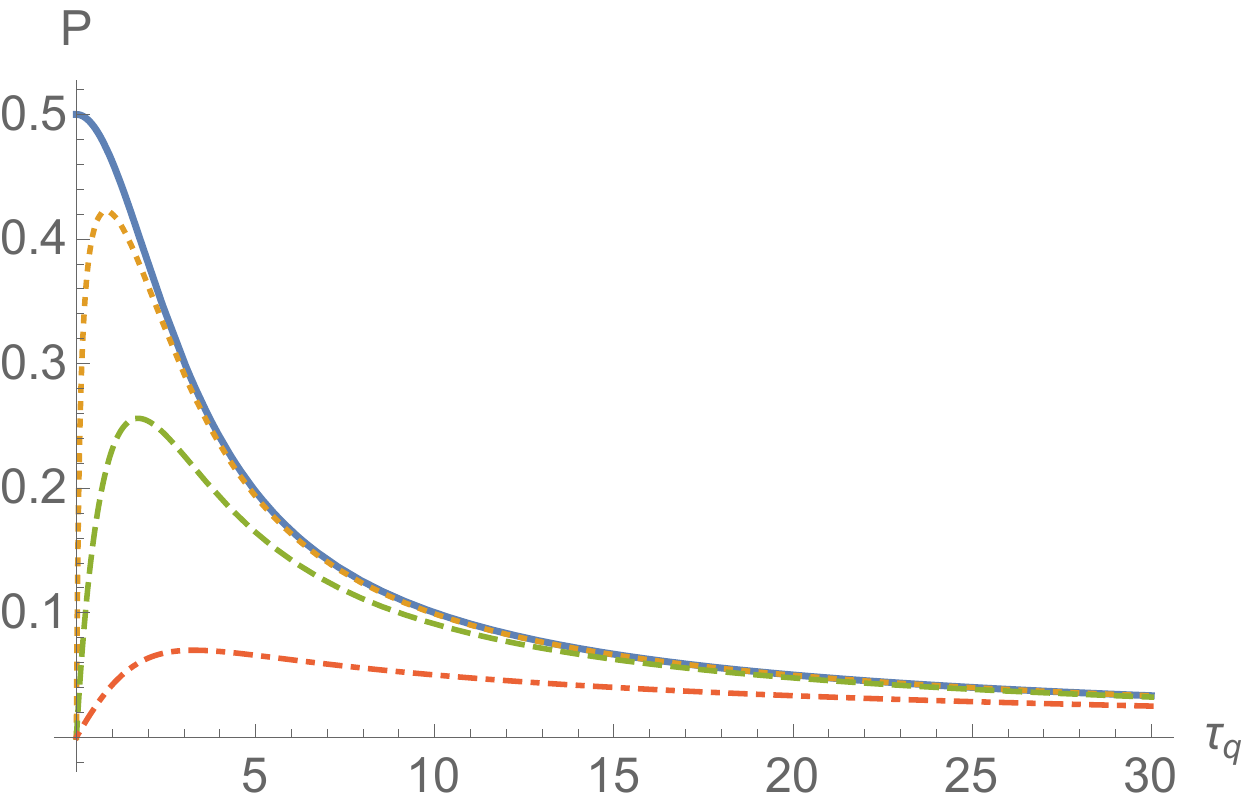}
  \caption{Power in $(\omega _A -\omega _B)(N_A-N_B)$ units versus
    thermalization stroke time $\tau _q$ with rates $\alpha _A = \alpha _B
    =1$ and swap  stroke times $\tau _w \rightarrow 0$ (solid line),
    $\tau _w=0.1$ (dotted), $\tau _w=1$ (dashed), and $\tau _w=10 $
    (dot-dashed). In all cases, the condition $\kappa \tau _w=\pi/2 $
    is supposed to hold.} 
\end{figure}
\par As in the two-mode bosonic Otto engine \cite{miobos}, short
thermalization times are expected to be harmful for the
signal-to-noise ratios of work and heat, with significant modification
in the thermodynamic uncertainty relations. As an example, in Fig. 8
we plot the signal-to-noise ratio for fixed value of the parameter
$N_B=2$ versus varying $N_A$ in the case of dimension $d=9$, for
different values of $\alpha \tau _q$, where it is apparent the
detrimental effect of decreasing the thermalization times. These
results can be obtained by replacing $ f(\beta _A \omega _A,\beta _B
\omega _B ,d)$ with $f(\beta _A^* \omega _A,\beta _B ^* \omega _B ,d)$
in Eq. (\ref{invw}).
\begin{figure}[ht]
  \includegraphics[scale=0.48]{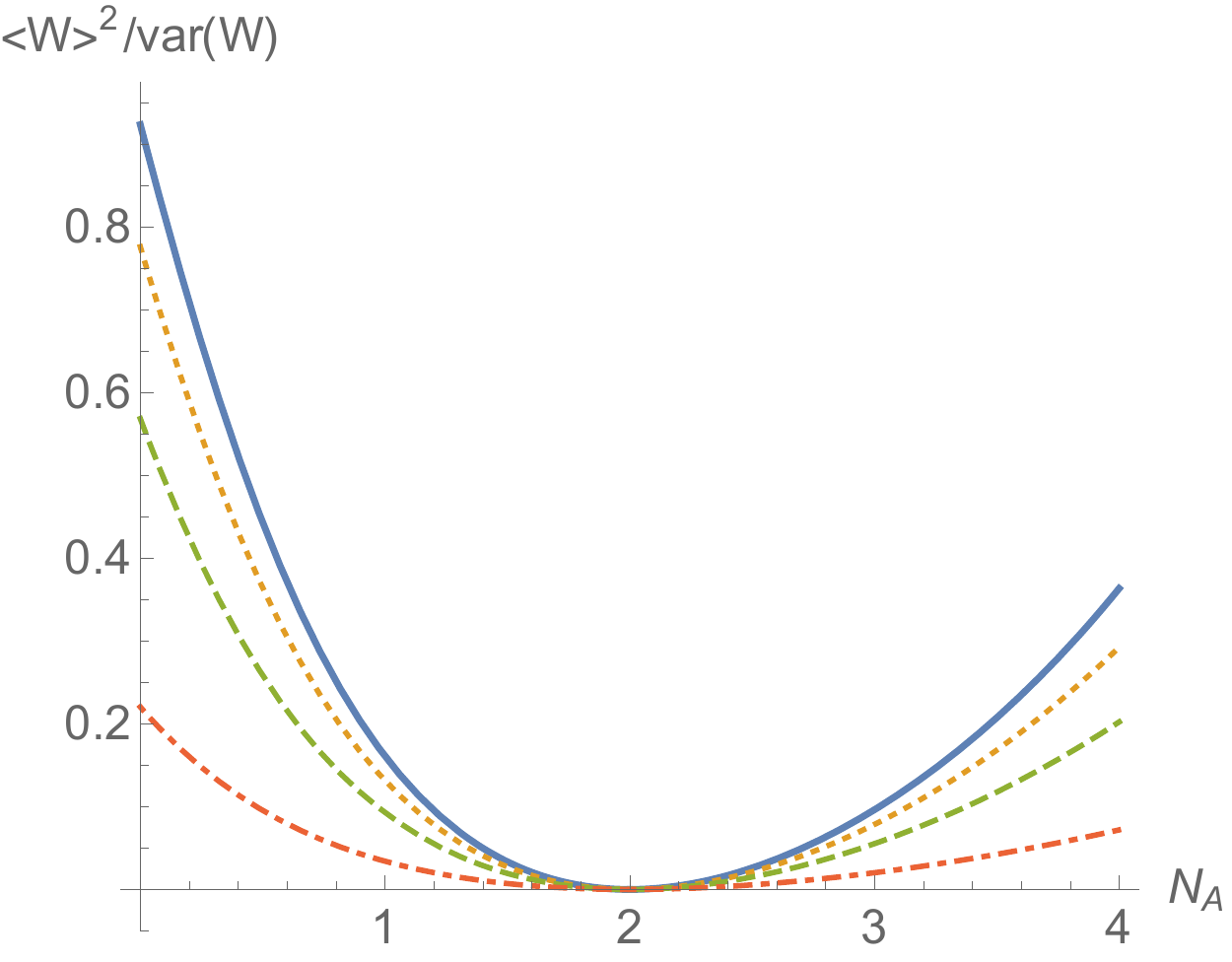}
\caption{Signal-to-noise ratio of the work for a 9-level swap engine ($\theta =\frac \pi
  2$) with mean occupation number $N_B=2$ versus varying $N_A$ for ideal
  thermalization (solid), and finite thermalization stroke times $\alpha \tau _q
  =3$ (dotted), $2$ (dashed), and $1$ (dot-dashed).}
\end{figure}
\par A deeper study of the finite-time scenario will be deferred for
future work, possibly by the explicit modeling of the thermal relaxation via master
equations and also by taking into account the residual correlations
between the qudits in the steady state of the limit cycle in the case
of partial-swap strokes.

\section{Conclusions}
\par By adopting the two-point-measurement scheme for the joint
estimation of work and heat, we have quantified the work and heat
fluctuations pertaining to two-qudit quantum thermodynamic two-stroke
Otto engines, where the work is extracted or performed by a 
multilevel partial-swap interaction, and thermal relaxation to two respective
reservoirs at different temperatures guarantees the cyclicity of the
protocol.  Exact relations among work, heat, fluctuations, efficiency,
and reliability emerge.

We have derived the characteristic function for work and heat, and
obtained the full joint distribution of the stochastic work and
heat. In all ranges of coupling parameter and dimension we have shown
thermodynamic uncertainty relations that reveal the interdependence
among average extracted work, fluctuations and entropy production,
confirming that reducing the noise-to-signal ratio of work or heat
comes at a price of increased entropy production.

The link between fluctuation theorems and thermodynamic uncertainty
relations appears to be relevant for the design of quantum
thermodynamic machines. In fact, the presented results match the
two-qubit swap engine with the two-mode bosonic case. In particular,
we have shown that the small violation of the standard TUR for
qubits is rapidly washed out for increasing dimension of the working
qudits.

The effect of partial thermalization due to finite-time thermal
strokes has also been studied in the case of perfect-swap 
unitary interaction. This results in a non-zero output power engine
where the allocation of thermal and working strokes have been
optimized. Thus, the efficiency at maximum power can be evaluated, and
violations of the Curzon-Ahlborn limit are observed. Such violation
are stronger for qubits and decrease for increasing dimension. We
note, however, that partial thermalization sensibly decreases the
signal-to-noise ratio of work and heat.

\appendix
\section{Derivation of the characteristic function $\chi(\lambda
  ,\mu)$ in Eq. (\ref{chi1})}
The characteristic function is given by the Fourier transform of the
joint probability $p(W, Q_H)$ of the stochastic work $W$ and heat
$Q_H$,
namely
\begin{eqnarray}
\chi(\lambda , \mu )=\int dW \int dQ_H \,
p(W,Q_H)\, e^{i \lambda W + i \mu Q_H} \;. \label{tf}
\end{eqnarray}
Let us adopt a two-point measurement protocol, where both the
Hamiltonian $H_A$ and $H_B$ of the two isolated systems
are measured just before and after the action of the unitary
interaction $U(\tau _w)$.  The probability $p_{n,m}$ for the initial measurement
outcomes $n$ and $m$ is given by the Gibbs weight of the initial state
$\rho _0$, namely
\begin{eqnarray}
  p_{n,m}=\frac{1}{Z_A Z_B}
  e^{-\beta _A \omega _A n}e^{-\beta _B \omega
  _B m}\;.
\end{eqnarray}
The conditional probability $q(l,s|n,m)$ pertaining to the final measurement
outcomes $l$ and $s$ given initial outcomes $n$ and $m$ is given by
\begin{eqnarray}
q(l,s|n,m)= |\langle l |\otimes  \langle s| U(\tau _w)|n \rangle
\otimes |m \rangle |^2
  \;.
\end{eqnarray}
For this occurrence characterized by the set $\{n,m,l,s\}$
note that the work and heat are given by
$\Delta E_A +\Delta E_B =\omega _A (l-n)+ \omega _B (s-m)$ and
$-\Delta E_ A = \omega _A (n-l)$,
respectively. Hence, the joint probability $p(W,Q_H)$ is obtained by
the following average over all occurrences
\begin{eqnarray}
  p(W,Q_H)&&=\sum _{n,m,l,s} p_{n,m} \, q(l,s|n,m) \, \delta (W- \omega _A (l
  -n)- \omega _B (s-m)) \, \delta (Q_H - \omega _A (n-l))
\;.\label{pwqq}
\end{eqnarray}
By replacing Eq. (\ref{pwqq}) in (\ref{tf}) and applying the delta
functions one obtains 
\begin{eqnarray}
  \chi(\lambda , \mu )&&=\frac{1}{Z_A Z_B}
  \sum _{n,m,l,s}   e^{-\beta _A \omega _A n}e^{-\beta _B \omega _B m} 
  e^{i \lambda [\omega _A (l-n)+ \omega _B (s-m)]} e^{i\mu \omega _A
    (n-l)}\nonumber \\& & \times 
  \Tr [U^\dag (\tau _w) (|l \rangle \langle l|
    \otimes |s \rangle \langle s|) U(\tau _w) (|n \rangle \langle
    n| \otimes |m \rangle \langle m|)] \nonumber \\& &=
\Tr [U^\dag (\tau _w) (e^{i (\lambda -\mu) H_A}
  \otimes e^{i \lambda H_B})
    U (\tau _w) (e^{-i (\lambda - \mu )H_A}\otimes e^{- i \lambda H_B})
    \rho_0 ] \;.
\label{a5}
\end{eqnarray}

\section{Evaluation of the characteristic function $\chi(\lambda
  ,\mu)$ in Eq. (\ref{chif})}
%\appendix
For the unitary operator $U (\tau _w)= U_0 (\tau _w) V_\theta $, since
$U_0 (\tau _w )$ commutes with both $H_A$ and $H_B$, Eq. (\ref{chif}) rewrites as
\begin{eqnarray}
  \chi (\lambda ,\mu ) &&=
\frac{1}{Z_A Z_B}
\Tr [V_\theta ^\dag  (e^{i (\lambda - \mu )H_A} \otimes e^{i \lambda H_B})
    V_\theta  (e^{-[i(\lambda - \mu )+ \beta _A] H_A}
    \otimes e^{-(i \lambda + \beta _B) H_B})]\;.\label{b1}
\end{eqnarray}
Notice that the equivalence of the two expressions (\ref{chif}) and
(\ref{b1}) implies that by replacing the time-dependent interaction
Hamiltonian in Eq. (\ref{ht}) with a constant quench Hamiltonian
$\kappa E$ with strong parameter $\kappa \gg 1$ and short duration $\tau
_w \ll 1$ with finite $\theta = \kappa \tau_ w$ then one obtains the same
statistics for the work and heat.  
\par Since $V_\theta = \cos\theta I -i \sin \theta E$, from
Eq. (\ref{b1}) one has 
\begin{eqnarray}
  \chi (\lambda ,\mu )&&
  = \cos^2 \theta +
  \sin^2 \theta
  \frac{1}{Z_A Z_B}
\Tr [E (e^{i (\lambda - \mu )H_A} \otimes e^{i \lambda H_B}) E
    (e^{-[i(\lambda - \mu )+ \beta _A] H_A}
  \otimes e^{-(i \lambda + \beta _B) H_B})]\nonumber \\& &
+ i \sin \theta \cos \theta
  \frac{1}{Z_A Z_B}
\{\Tr [E (e^{i (\lambda - \mu )H_A} \otimes e^{i \lambda H_B}) 
    (e^{-[i(\lambda - \mu )+ \beta _A] H_A}
    \otimes e^{-(i \lambda + \beta _B) H_B})]
\nonumber \\& &   -\Tr [ (e^{i (\lambda - \mu )H_A} \otimes e^{i \lambda H_B}) E
    (e^{-[i(\lambda - \mu )+ \beta _A] H_A}
  \otimes e^{-(i \lambda + \beta _B) H_B})]\}
\nonumber \\& &= 
\cos^2 \theta +
  \sin^2 \theta
 \frac{1}{Z_A Z_B}
\Tr [e^{(i \lambda \frac {\omega _B -\omega _A}{\omega _A} + i\mu -
    \beta _A)H_A} \otimes
  e^{-(i \lambda \frac {\omega _B -\omega _A}{\omega _B} + i\mu \frac
    {\omega _A}{\omega _B}+
    \beta _B)H_B}]\;,\label{chif9}
\end{eqnarray}
where we used the identity $E (f(H_A)\otimes g(H_B))E =g (\omega _B H_A
/\omega _A) \otimes f (\omega _A H_B/ \omega _B)$ that holds for
arbitrary functions $f$ and $g$ to simplify the factor of $\sin ^2
\theta $, and we applied the property $\Tr[E
  (X\otimes Y)]=\Tr[ (X\otimes Y)E]= \Tr [X Y]$ for all operators $X$
and $Y$ to cancel out  the two terms that multiply $\sin\theta \cos
\theta $. Finally, by evaluating the trace we
obtain 
\begin{eqnarray}
  \chi (\lambda ,\mu )&&=
\cos^2 \theta + \sin^2 \theta
\frac{1}{Z_A Z_B} \frac{1-e^{-d (\beta _A \omega _A + i \xi)}}
     {1-e^{- (\beta _A \omega _A + i \xi)}}
      \frac{1-e^{-d (\beta _B \omega _B - i \xi)}}
          {1-e^{- (\beta _B \omega _B - i \xi)}} \nonumber \\& &
= \cos^2 \theta + \sin^2 \theta \frac{\sinh \left ( \frac{\beta _A
    \omega _A}{2}\right )\sinh \left ( \frac{\beta _B
    \omega _B}{2}\right )\sinh \left [ \frac d2 (\beta _A
    \omega _A + i \xi  ) \right ]\sinh \left [ \frac d2 (\beta _B
    \omega _B - i \xi  ) \right ]}
{\sinh \left ( \frac{ d \beta _A
    \omega _A}{2}\right )\sinh \left (  \frac{d \beta _B
    \omega _B}{2}\right )\sinh \left [ \frac 12 (\beta _A
    \omega _A + i \xi  ) \right ]\sinh \left [ \frac 12 (\beta _B
    \omega _B - i \xi  ) \right ]}
  \;,\label{chif2}
\end{eqnarray}
where $\xi = (\omega _A -\omega _B) \lambda -  \omega _A \mu$.

\section{Efficiency at maximum work per cycle for $d\rightarrow \infty $}
For $d\rightarrow \infty $ the maximum work per cycle is obtained
from Eqs. (\ref{gx}) and (\ref{wco}) with $\theta =\pi/2$, and one has
\begin{eqnarray}
  |\langle W \rangle |
  =\frac {\omega _A -\omega _B}{2}\left [ \coth
  \left ( \frac{\beta _A \omega _A}{2}\right ) -
  \coth
  \left ( \frac{\beta _B \omega _B}{2}\right )
  \right ]\;.
\end{eqnarray}
For fixed ratios $\eta _C =1- \frac {T_B}{T_A}
$ and $\eta=1- \frac{\omega
  _B}{\omega _A}$ one can rewrite
\begin{eqnarray}
  |\langle W \rangle |=T_B \frac{\eta }{1-\eta } x 
\left [ \coth
  \left (  \frac{1-\eta _C}{1-\eta }x \right ) -
  \coth
  \left ( x\right )
  \right ]\;,
\end{eqnarray}
with $x=\frac{\beta _B \omega _B}{2}$. Along similar lines as in
Appendix D of Ref. \cite{picc} one easily shows that $|\langle W
\rangle |$ achieves the maximum for $x\rightarrow 0$, since $\partial
_x |\langle W \rangle | <0$ for all $x>0$. Hence, the maximum can be
searched in the high-temperature limit \cite{also}, where 
\begin{eqnarray}
|\langle W \rangle | \simeq T_B \frac{\eta (\eta _C -\eta )}{(1-\eta
  _C)(1-\eta )}=T_B \left (1 - \frac{\omega _A}{\omega _B} \right )
+ T_A \left (1 - \frac{\omega _B}{\omega _A} \right )
\;,
\end{eqnarray}
which is maximized by the Curzon-Ahlborn value $\eta _m =\eta
_{CA}=1-\sqrt{T_B/T_A}$.


\begin{thebibliography}{99}
\bibitem{nota}This can be easily inspected from 
  Eq. (\ref{chif}), since the replacements $\lambda \rightarrow i \beta _B
  - \lambda $ and $\mu \rightarrow i(\beta _B -\beta _A)-\mu$ are
  equivalent to $ \beta _A \omega _A + i\xi \leftrightarrow \beta _B
  \omega _B - i\xi $.
\bibitem{notafc}This can be explicitly seen by comparing the
  expression for the characteristic function $\chi(\lambda ,\mu)$ in
  Eq. (\ref{b1}) with Eq. (6) of Ref. \cite{miobos} when in both
  equations one takes $\theta =\pi/2$.
\end{thebibliography}
\end{document}